\documentclass[preprint]{aastex}
\usepackage{mkfig}

\begin{document}

\title{Testing the Flux Rope Paradigm for Coronal Mass Ejections Using
  a Three Spacecraft Encounter Event} 

\author{Brian E. Wood\altaffilmark{1}, Phillip Hess\altaffilmark{1}}
\altaffiltext{1}{Naval Research Laboratory, Space Science Division,
  Washington, DC 20375, USA; brian.e.wood26.civ@us.navy.mil}


\begin{abstract}

     We present a 3-D morphological and field reconstruction of a
coronal mass ejection (CME) from 2023~November~28, which hits three
spacecraft near 1~au: Wind at Earth's L1 Lagrange point;
STEREO-A with a longitudinal separation of $6.5^{\circ}$ west of
Earth; and Solar Orbiter (SolO) at $10.7^{\circ}$ east of Earth.  The
reconstruction assumes a magnetic flux rope (MFR) structure for the
CME.  With this event, we test whether field tracings observed by
a spacecraft passing near the central axis of a CME MFR (STEREO-A)
can be used to successfullly predict the field behavior seen by
a spacecraft $17^{\circ}$ away (SolO), which has a more grazing
encounter with the CME.  We find that the MFR model does have
significant success in simultaneously reproducing the field signs
and rotations seen at STEREO-A, Wind, and SolO.  This provides
support for the MFR paradigm for CME structure.  However, the 
SolO measurements, which are farthest from the central
axis of the MFR, show less defined MFR signatures,
presumably due to a greater degree of erosion and degradation
of the MFR structure far from its central axis.

\end{abstract}


\section{Introduction}

     Future advancement in the understanding of coronal mass ejection
(CME) structure and interplanetary evolution is likely to rely on
observations that probe CME field structure in multiple locations.
Traditional studies that analyze magnetic field tracings through
a CME made by a single spacecraft have been invaluable in illuminating
the general characteristics of field structure, guiding the
development of the first models of global CME structure.  In particular,
the predominant magnetic flux rope (MFR) paradigm describes CMEs as
tubular structures permeated by helical field, with both legs of
the tube stretching back to the
Sun \citep{rpl90,cjf95,jc97,vb98,av13,bew17}.  However, proper testing
of the MFR paradigm requires measurements of CME properties in multiple
locations to see whether the properties of an MFR inferred locally truly
apply to the global CME.

     Past multipoint studies have mostly involved CMEs that hit spacecraft
operating at different distances from the Sun, beginning with events
observed by the Helios mission that probed the interplanetary
medium as near as 0.3~au from the Sun \citep{lb81}, combined
with observations made by spacecraft near Earth.
Alignments between spacecraft operating at 1~au (e.g., Wind, ACE,
STEREO) and spacecraft visiting other planets have provided other
opportunities \citep{rmw16,ow17,col18,swg18,nl20,ep21}.  There
are enough events of this nature that catalogs of them have been
developed, allowing for a more statistical analysis of how CME properties
evolve with radial expansion and propagation
\citep{bv19,swg19,tms20,cs22}.  The
launches of Parker Solar Probe (PSP) in 2018 and Solar Orbiter (SolO)
in 2020 provide new opporunties for multipoint CME studies
\citep[e.g.,][]{rmw21}, with both PSP and SolO operating inside 1~au.
\citet{cm22} have already provided a catalog of multipoint events
involving PSP and SolO, which is updated to the present
year (\url{https://www.helioforecast.space/lineups}).

     Sampling a CME's structure at different distances from the Sun has the
disadvantage of potentially confusing spatial variations in CME structure
with the effects of evolutionary changes as a CME moves away from the Sun.
A focus on spatial variation ideally involves multi-site sampling of a
CME at the same time.  An ideal opportunity for such studies has recently
occurred when STEREO-A passed by Earth in 2023~August.  Lauched in
2006~October, the twin STEREO spacecraft have long provided stereoscopic
imaging capabilities for CMEs erupting from the Sun, with one spacecraft,
STEREO-A, drifting ahead of Earth in its orbit around the Sun, and one
spacecraft, STEREO-B, drifting behind.  With an angular drift from Earth of
about $22^{\circ}$ per year, the two spacecraft eventually
behind the Sun in 2014-2015, when contact
with the two spacecraft was lost.  STEREO-A operations were successfully
resumed when it emerged from behind the Sun, but such was not the case for
STEREO-B.  Nevertheless, STEREO-A has continued observations since then, with
the spacecraft gradually drifting back toward Earth.

     From 2022~July until 2024~September, STEREO-A was within $25^{\circ}$
longitude of Earth, an ideal time to look for CMEs that happen to hit both
STEREO-A and spacecraft operating at the Earth L1 point (e.g., Wind).
The two STEREO spacecraft were also near Earth shortly after the 2006 launch
date, but this was in a very inactive period for the Sun, greatly limiting
the number of CMEs that hit the spacecraft while near Earth
\citep{cjf11}.  In contrast, solar activity was quite high during
the more recent passage of STEREO-A past the Sun.  Nevertheless, from
2020~October until 2022~August, when STEREO-A moved from $60^{\circ}$ to
$20^{\circ}$ from the Sun-Earth line, \citet{nl24}
found only four events
that hit both STEREO-A and Wind near Earth, and none of those four provided
clean MFR signatures at both locations that would allow inferences of MFR
orientation at the two sites.  This suggests that the angular extent of CMEs
in this period have generally been significantly smaller than generally
assumed, with typical angular widths of only $\sim 20^{\circ}-30^{\circ}$.

     We here provide an analysis of an event that hit both STEREO-A and
Wind at a significantly smaller angular separation of only $6.5^{\circ}$.
This CME, which erupted from the Sun on 2023~November~28,
not only hits STEREO-A and Wind but also hits a third spacecraft, SolO,
which happened to be near Earth at the time in both distance and longitude.
More specifically, SolO was 0.84~au from the Sun, and $10.7^{\circ}$
away from Earth in longitude, in the opposite direction from STEREO-A.
The event has previously been studied
by \citet{zs24} and \citet{yc24}.
With three spacecraft tracks through the CME near 1~au, this is an ideal
event for testing the MFR paradigm.  We therefore provide a full
3-D reconstruction of the CME morphology based on all available white light
imaging of the event; and we then assess whether a reasonable MFR field
model can be inserted into the inferred MFR shape that can reproduce the
field observations from all three spacecraft.

\section{Images of the CME}

\begin{figure}[t]
\plotfiddle{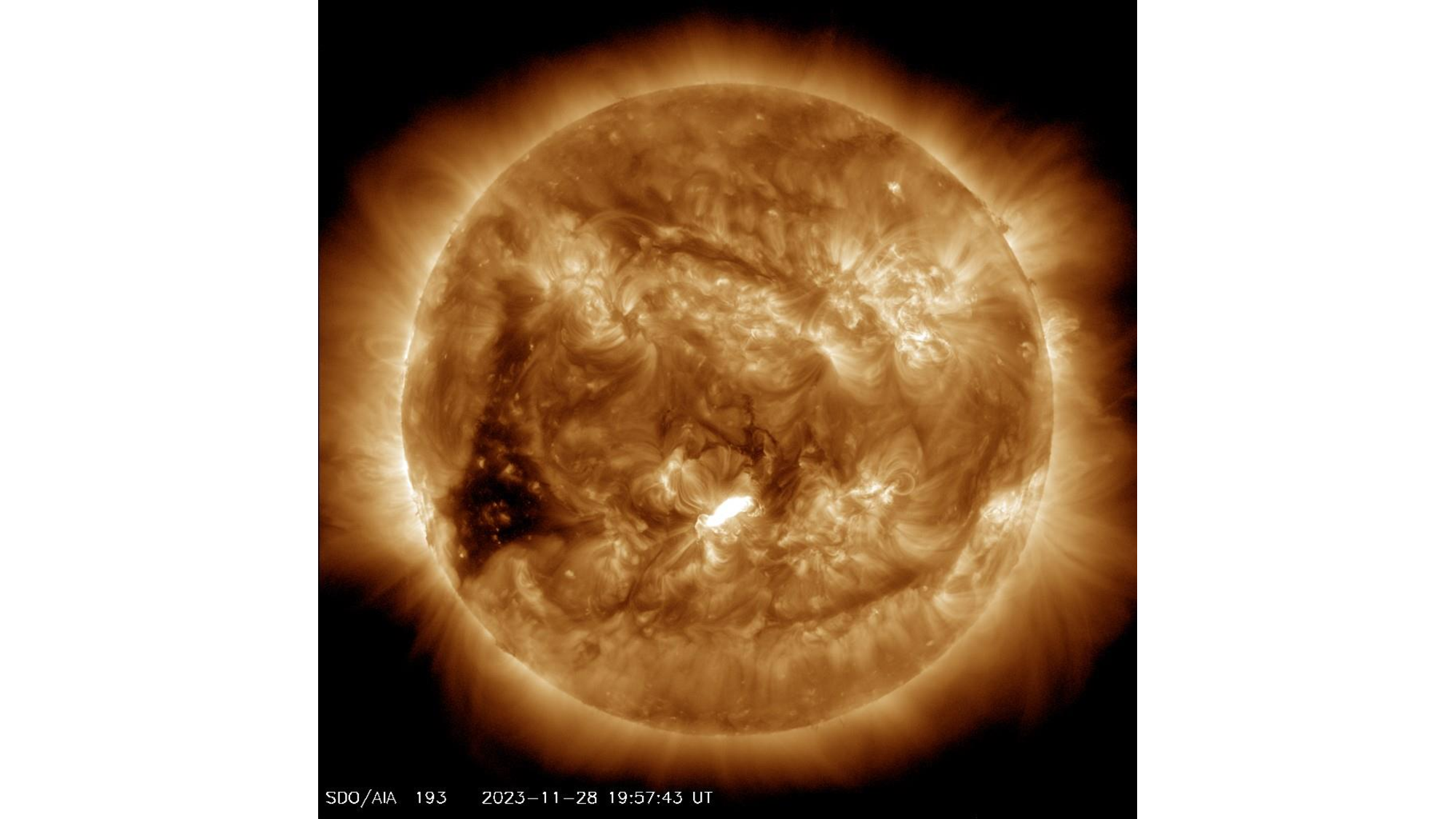}{2.5in}{0}{40}{40}{-190}{-10}
\caption{SDO/AIA 193~\AA\ channel image showing the M9.8 flare near
  disk-center that accompanies the 2023~November~28 CME.  Also of note
  is the dark coronal hole east of the flare site.}
\end{figure}
     At about UT 19:17:53 on 2023~November~28, an M9.8 flare began on the
Sun, near disk center as viewed from Earth.  Figure~1 shows an EUV image of
this flare from the 193~\AA\ channel of the Atmospheric Imaging Assembly
(AIA) instrument on board the Solar Dynamics Observatory (SDO).
\citet{zs24} provide a detailed description of the flare morphology.
They decompose the surface activity into three separate erupting MFRs,
though it should be noted that it is unclear how these little MFRs
relate to the big CME-sized MFR that we are modeling here.  The flare
was accompanied by a halo CME observed by the C2 and C3 coronagraph
constituents of the Large Angle Spectrometric Coronagraph (LASCO)
instrument on board the Solar and Heliospheric Observatory (SOHO), which
has been operating at Earth's L1 Lagrangian point since 1996
\citep{geb95}.  The CME would hit Earth on 2023~December~1,
resulting in a modest $K_p=7$ geomagnetic storm.

\begin{figure}[t]
\plotfiddle{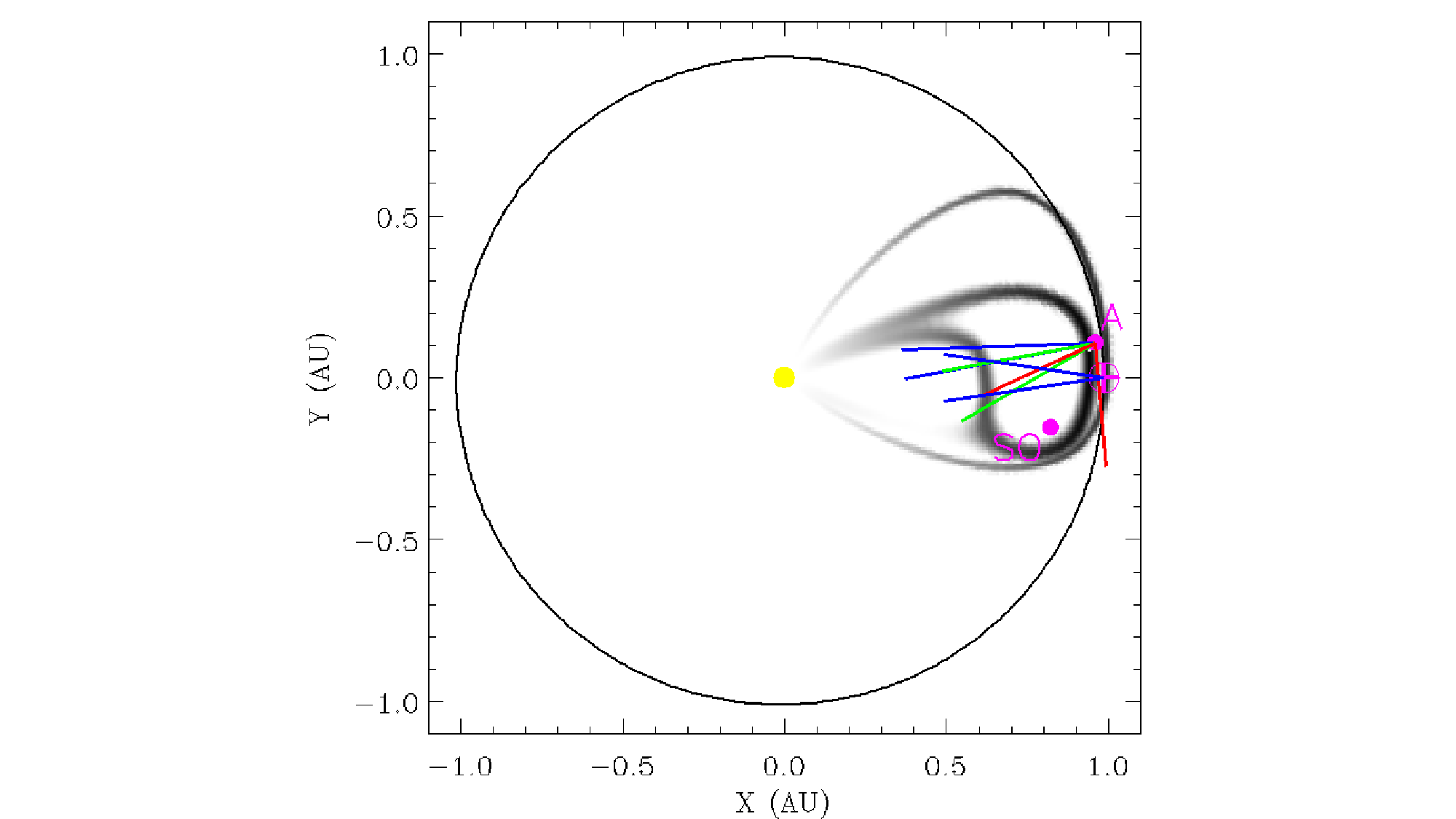}{2.5in}{0}{40}{40}{-200}{-10}
\caption{Ecliptic plane map for 2023~December~1, in an HEE coordinate
  system, showing the positions of STEREO-A (``A'') and Solar Orbiter
  (``SO'') relative to Earth.  Blue lines emanating from the Earth's
  location indicate the field of view of the LASCO/C3 coronagraph.
  Blue, green, and red lines from STEREO-A's location indicate the
  fields of view of COR2-A, HI1-A, and HI2-A.  The grayscale image
  indicates an ecliptic plane slice through the 3-D reconstruction of
  the 2023~November~28 CME shown in Figure~6.  The lower
  leg of the MFR is below the ecliptic and therefore not apparent.}
\end{figure}
     We here perform a full 3-D reconstruction of the morphology of this
CME based on the available white light imaging of this event.  This
includes the aforementioned SOHO/LASCO C2 and C3 images, covering
plane-of-sky distances from Sun-center of
1.5--6 R$_{\odot}$ and 3.7--30 R$_{\odot}$, respectively.
The event is also observed by the imagers on board STEREO-A.
These include the COR2-A coronagraph, observing at angular distances from
Sun-center of $0.7^{\circ}-4.2^{\circ}$ ($2.5-15.6$ R$_{\odot}$), and also the
HI1-A and HI2-A heliospheric imagers, viewing at $3.9^{\circ}-24.1^{\circ}$
and $19^{\circ}-89^{\circ}$, respectively \citep{rah08,cje09}.
Figure~2 shows the location of STEREO-A in the
ecliptic plane relative to Earth and SOHO, in a heliocentric Earth
ecliptic (HEE) coordinate system, and also shows the fields of
view of the various imagers used here.  STEREO-A is only $6.5^{\circ}$
longitude ahead of Earth at the time of the event.  Also
shown is an ecliptic plane slice through the 3-D reconstruction that
is described in Section~4.

\begin{figure}[t]
\plotfiddle{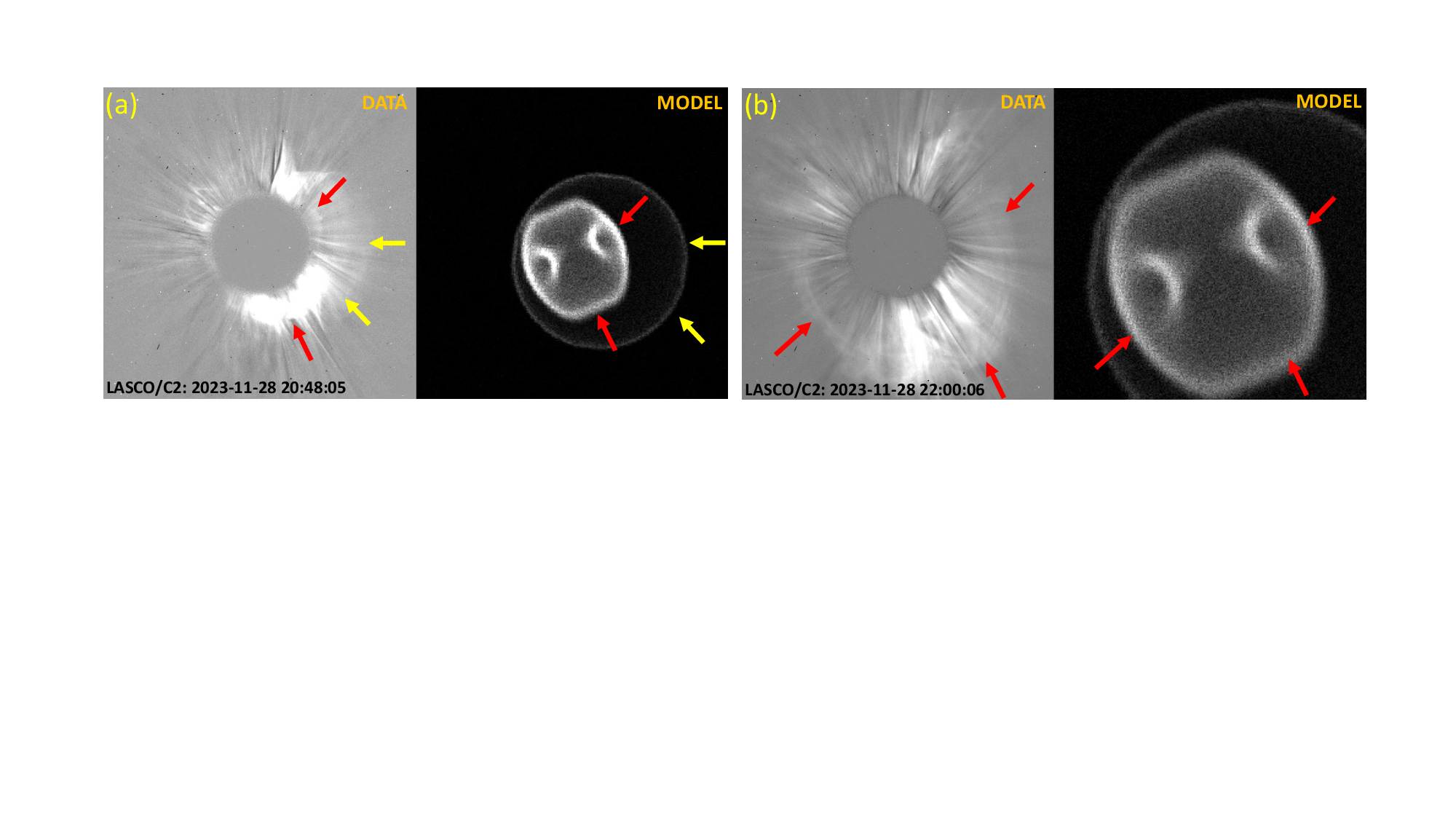}{1.6in}{0}{55}{55}{-265}{-160}
\caption{(a) The left panel is a LASCO/C2 image of the 2023~November~28
  CME from UT 20:48:05, which can be compared with the synthetic image
  shown in the right panel, based on the 3-D reconstruction of the event
  from Figure~6.  Red arrows mark the presumed leading edge of the CME
  ejecta, while yellow arrows mark the leading edge of a shock
  created by the fast CME, which is only clearly visible west and north
  of the ejecta.  (b) Similar to (a), but for a LASCO/C2 image from
  UT 22:00:06.  A movie associated with this figure provides a more
  comprehensive presentation of the LASCO data/model comparison, for
  both LASCO/C2 and LASCO/C3 images.}
\end{figure}
\begin{figure}[t]
\plotfiddle{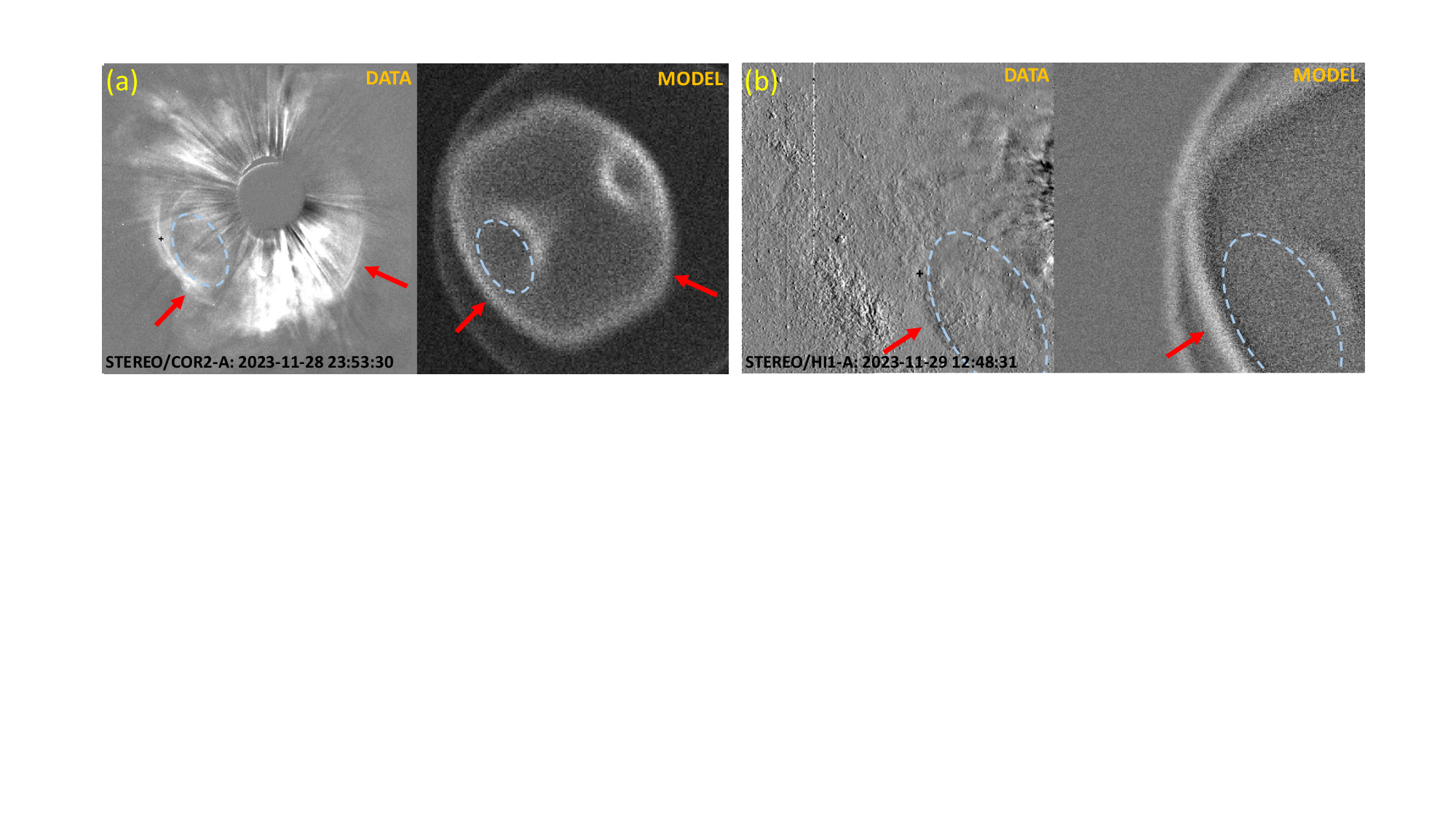}{1.6in}{0}{55}{55}{-265}{-160}
\caption{(a) The left panel is a STEREO/COR2-A coronagraphic image of
  the 2023~November~28 CME, which can be compared with the
  synthetic image shown in the right panel, based on the 3-D
  reconstruction of the event from Figure~6.  Red arrows mark the
  presumed leading edge of the CME ejecta.  A light blue ellipse
  outlines the top of the eastern leg of the MFR structure,
  apparent in both the real and synthetic images.  (b) Similar to
  (a), but for a later STEREO/HI1-A image.  A movie associated with this
  figure provides a more comprehensive presentation of the STEREO-A
  data/model comparison, including COR2-A, HI1-A, and HI2-A images.}
\end{figure}
     Figures~3-4 show a selection of the LASCO and STEREO-A images of the
CME that are the basis of our 3-D reconstruction.  A more thorough
presentation of the imagery is provided in two movies associated with
these figures.  This is a halo CME
from the perspectives of both SOHO/LASCO and STEREO-A.  The CME is fast
enough that there is a clear shock ahead of the bright ejecta, but the
shock is only clearly visible to the west and north of the ejecta (see
Figure~3(a)).  Even though SOHO and STEREO-A are only $6.5^{\circ}$ apart
in longitude, comparison of Figure~3(b) and Figure~4(a) shows that there
is still a very clear shift in the CME halo position relative to the Sun.
The halo is more centered on the Sun as viewed by SOHO/LASCO, while for
COR2-A the halo center is shifted to the left of the Sun.  Our
determination of the CME's central trajectory direction promises to be
extremely precise due to the diagnostic power of this clear shift.

     If SOHO/LASCO and STEREO-A were instead viewing this event from
a lateral perspective, there might be little difference in the CME's\
appearance in LASCO/C2 and COR2-A images, with a spacecraft separation
of only $6.5^{\circ}$, and therefore little stereoscopic diagnostic power.
But with this CME in fact directed right at SOHO/LASCO and STEREO-A, the
situation is dramatically different.  There have been many analyses trying
to estimate uncertainties in CME trajectory directions from stereoscopic
analysis \citep[e.g.,][]{cv23}, but the CME studied here
illustrates just how difficult it is to provide uncertainty
estimates for stereoscopic analysis when it can be so dependent on the
exact viewing geometry.

     In Section~4, we will be presenting a 3-D morphological reconstruction
of the CME, assuming an MFR shape, with synthetic model images from this
reconstruction shown in Figures~3-4.  An attractive feature of
this particular event is that we believe there is evidence for an MFR
shape in the white light imaging data, particularly from STEREO-A.
For a halo CME viewed from the front, the leading edge
of the MFR will define the extent of the halo, while
the legs of the MFR, if visible at all, should be seen
as roughly circular structures on the sides of the halo, as suggested by
the synthetic coronagraphic images in Figures~3-4.  We believe we see
this in the COR2-A and HI1-A images, where we see an elliptical outline
that we attribute to the top of the east leg of the MFR.
Light blue ellipses are used
to identify this structure in Figure~4.  Structures such as this are
not apparent in all halo CMEs.  It is unclear what this means for
the MFR paradigm, as the visibility of the MFR legs will depend on
whether there is sufficient mass loaded onto the sides of the MFR
legs to visibly outline them in the images.  For the 2023~November~28
CME, we do see something we can interpret as the outline of an MFR leg,
but in other cases the mass loading may not provide such an outline.

\section{Kinematic Modeling}

     In the previous section we noted that the viewing geometry of this
event promised to provide a particularly accurate determination of CME
trajectory direction.  In contrast, this viewing geometry is lousy
for measuring CME kinematics.  There is no available viewing platform
that provides a lateral perspective of this event, which would allow
us to properly see the CME leading edge expanding radially away from the
Sun, yielding a direct indication of CME speed.  Instead, the
CME halo leading edges offered by SOHO/LASCO's and STEREO-A's frontal
perspectives are indicative of lateral expansion of the CME, which we
can only hope provides some constraint on the radial expansion rate.

     We here use the STEREO-A images to estimate a kinematic model for
the CME, where we track the more extended east side of the halo to
measure leading edge elongation angle ($\epsilon$) as a function of
time, based on the COR2-A, HI1-A, and HI2-A images.  Converting
$\epsilon$ to physical distance from Sun-center ($r$) requires a
geometric approximation for the shape of the CME.  The two most
commonly used assumptions are the ``Fixed-$\phi$'' (FP) approximation and
the ``Harmonic Mean'' (HM) approximation.  The former assumes the CME is
infinitely narrow, leading to the relation
\begin{equation}
r=\frac{d\sin \epsilon}{\sin(\epsilon+\phi)},
\end{equation}
where $d$ is the distance from the observer to the Sun and $\phi$ is
the angle between the CME trajectory and the observer's line of sight
to the Sun \citep{swk07,nrs08}.
The latter approximates the CME as a sphere centered halfway
between the Sun and the CME leading edge \citep{nl09}, where
\begin{equation}
r=\frac{2d\sin \epsilon}{1+\sin(\epsilon+\phi)}.
\end{equation}
The FP relation will only work for narrow CMEs where the
apparent leading edge from the relevant perspective is tracking a
part of the CME front that is close to the actual CME leading edge.
This is very far from the case for a halo event, so the FP
approximation is not a possibility here.  The HM
approximation should be better, but despite the success we have
had in using this approximation in the past for a wide variety of
events \citep{bew17}, we find that it fails badly here.

     In the next section we will be presenting a 3-D reconstruction of
the CME, and synthetic images derived from it.  The kinematic model
we are deriving in this section is used to describe how the 3-D
model MFR CME expands with time.  A successful kinematic model
should result in synthetic images that have the apparent leading
edge close to the observed locations, as quantified by the $\epsilon$
values that are the starting point of the kinematic analysis.
It is in this sense that the HM approximation
clearly fails badly.  The problem is that the HM approximation
assumes a CME that is simply much too broad compared to the actual
CME, which is a problem here since the apparent leading edges that we
are following for our halo event are more indicative of the CME's lateral
extent than its radial distance from the Sun.  We could in principle
derive a kinematic model directly from the parametrized MFR shape
described in the next section, but because there is no simple equation
connecting $\epsilon$ and $r$ using the shape parameters, it would be
necessary to go image by image to infer the expansion factors
for the shape as a function of time that yield the correct leading
edge distance.  Furthermore, this tedious process would have to be
repeated every time the MFR shape parameters are changed, which is
impractical for the trial-and-error process of converging on a
best-fit MFR shape.  Thus, the use of a geometric approximation for
the kinematic analysis is required.

     For our kinematic model, we need a geometric approximation that
assumes a narrower CME, but not as narrow as the infinitely narrow
approximation represented by FP.  This leads us to
the ``Self-Similar Expansion'' (SSE) approximation championed by
\citet{jad12}.  Like HM, this approximation assumes
a spherical CME, but allows it to be centered anywhere between the Sun
and the CME leading edge, with an extra free parameter, $\lambda$,
indicating the angular half-width of the sphere as seen from the Sun.
This leads to
\begin{equation}
r=d\sin\epsilon \left[
  \frac{1+\sin\lambda}{\sin(\epsilon+\phi)+\sin\lambda} \right].
\end{equation}
The HM equation is recovered for $\lambda=90^{\circ}$, with lower $\lambda$
values corresponding to narrower CMEs.  Through experimentation we end up
assuming $\lambda=15^{\circ}$, with this seeming to resolve the fundamental
difficulty with HM noted above where the synthetic images of
the model MFR showed leading edges offset from the observed locations.
The top panel of Figure~5 shows the
resulting leading edge distances as a function of time.

\begin{figure}[t]
\plotfiddle{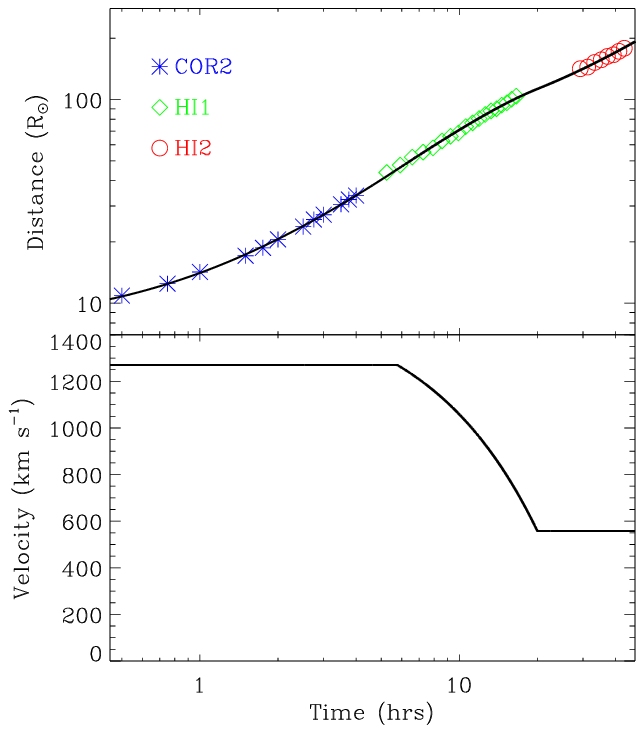}{3.0in}{0}{75}{75}{-230}{-285}
\caption{The top panel shows leading edge distance from Sun-center as a
  function of time for the 2023~November~28 CME, based on STEREO-A
  measurements from the COR2-A, HI1-A, and HI2-A imagers.  The measurements
  are fitted with a simple three-phase kinematic model, with a phase
  of constant velocity, followed by one of constant acceleration,
  and finally another phase of constant velocity.  This leads to the
  solid line fitted to the data points, with the bottom panel showing
  the inferred leading edge velocity as a function of time.  The
  time axis is relative to the $t=0$ time of UT~20:53:30
  on 2023~November~28.}
\end{figure}
     For our purposes, we require a kinematic model that describes the
CME's distance from the Sun at all times, from initiation to beyond
1~au if necessary.  We therefore fit the distance versus time data
in Figure~5 with a simple multi-phase model, analogous to what we
have done in the past \citep[e.g.,][]{bew17}.  The three-phase model
used here assumes an initial phase of constant velocity, followed by
a phase of constant acceleration, and then a final phase with constant
velocity.  There are only five free parameters of this simple model:
initial height, initial velocity, start time of phase~2, acceleration
during phase~2, and end time of phase~2.
The best fit is shown in the top panel of Figure~5, with
the bottom panel showing the inferred velocity profile.
The kinematic model suggests that the CME is at its peak
speed of 1270 km~s$^{-1}$ when it enters the COR2-A field
of view.  It decelerates during its journey through interplanetary
space, mostly while it is in the HI1-A field of view, reaching a
final velocity of 558 km~s$^{-1}$.  Our peak speed estimate
is in reasonably good agreement with the 1374 km~s$^{-1}$ estimate
of \citet{yc24}.

     In Section~5, we will be presenting the in~situ observations of
this CME, from Wind, STEREO-A, and SolO.  The kinematic model in
Figure~5 does not do a very good job of predicting the arrival time
of the CME at these spacecraft.  Although the use of
the SSE approximation has at least allowed a plausible kinematic
model to be inferred from the $\epsilon$ measurements, the model
is still clearly imprecise.

     Besides the aforementioned problem with the frontal viewing
geometry, another difficulty that should be mentioned is the dubious
nature of the HI2-A measurements in Figure~5.  The eastern
leading edge of the CME is easy to follow in COR2-A
and early in HI1-A, but the leading edge fades significantly as it
moves to the left side of the HI1-A field of view, where the presence
of the Milky Way (see Figure~4b) makes it impossible to follow.
This is the reason for the time gap between the HI1-A and HI2-A data
points in Figure~5.  Conclusively identifying the CME front in HI2-A
beyond the Milky Way is hard.  Inspection of the HI2-A movie
(associated with Figure~4) shows many fairly bright fronts that are
very tempting to associate with our CME.  However, the timing and
location of the fronts are wrong, and we believe that these fronts
are background corotating interaction region (CIR) fronts associated
with the high speed streams emanating from the dark coronal hole seen
to the east of the flare in Figure~1.  We do think we see hints of
our CME front in the HI2-A movie, but the $\epsilon$ measurements
from HI2-A should be considered very uncertain, yet another
reason for imprecision in the kinematic model.

     For an event propagating toward the spacecraft,
for the majority of the HI2-A field of view the scattering
efficiency will be low and the CME can be overshadowed
by brighter background features.  This could also be in part why we
are unable to identify the CME in any images from the SoloHI
heliospheric imager \citep{rah20}, despite SolO being hit
by the CME.  SoloHI has previously been used in multiviewpoint CME
studies \citep{ph23}, but in this case it is looking in the
wrong direction, away from Sun-Earth line where most of the CME is
from SolO's perspective.

\section{Morphological Reconstruction}

     We model the 3-D morphology of the 2023~November~28 CME
using a methodology utilized many times before in stereoscopic image
analysis.  Previous works describe how parametrized MFR shapes and
shock shapes can be constructed \citep[e.g.,][]{bew09,bew17,bew21}.
Density cubes containing
these structures are created, with mass placed on the
surfaces to indicate the outline of the 3-D shape.  No attempt
is generally made to model internal MFR structure, though there have
been exceptions \citep{bew21}.  From the
density cubes, synthetic images of the model CME can be made for
comparison with the actual white light images of the event.
Simple self-similar expansion is assumed for the CME, meaning that
the shape does not change, with the kinematic model in Figure~5
defining how the scale size of the 3-D model increases with time.

\begin{figure}[t]
\plotfiddle{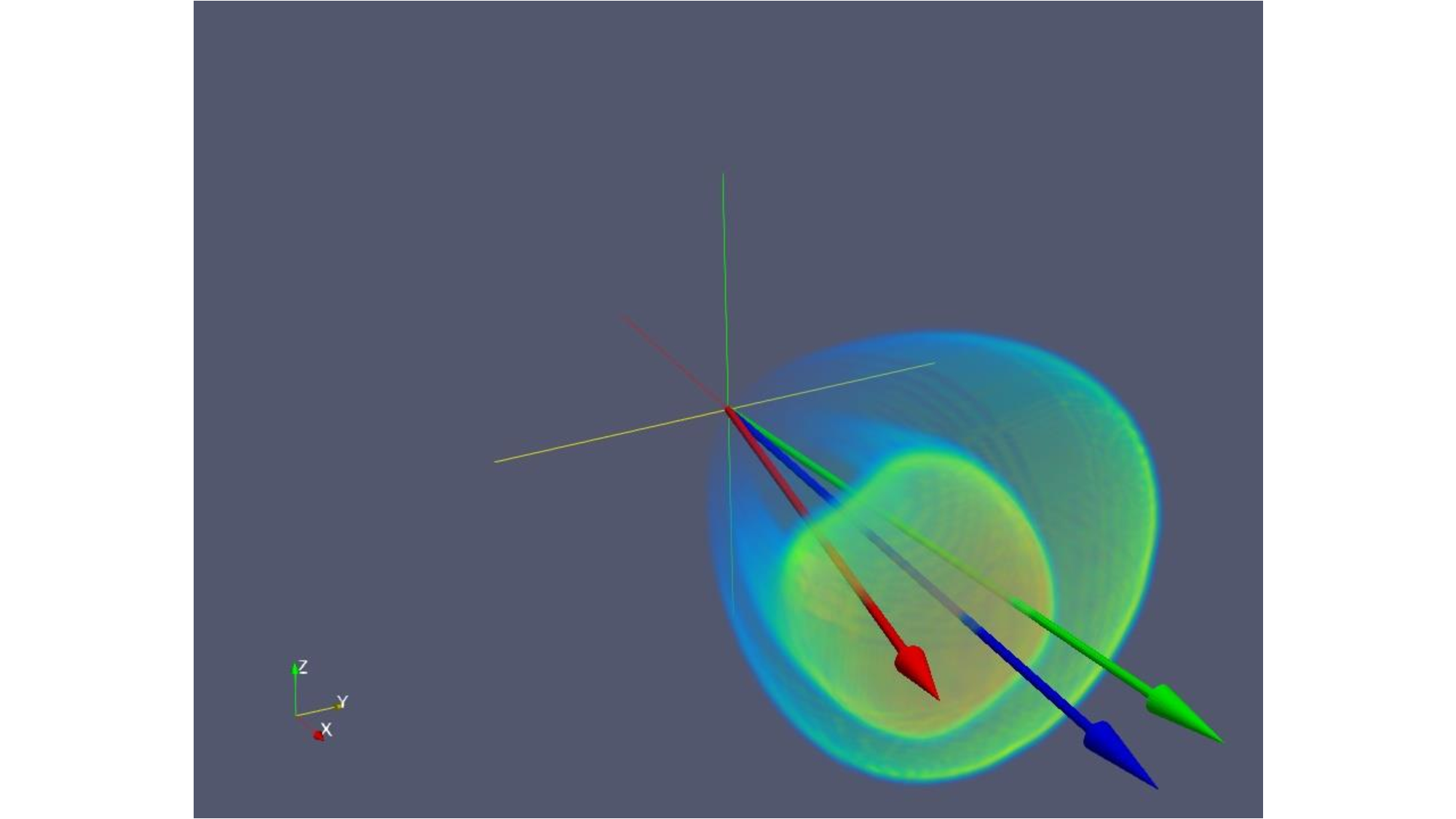}{3.0in}{0}{45}{45}{-210}{-10}
\caption{Reconstructed 3-D structure of the 2023~November~28 CME,
  consisting of an MFR shape and a lobular shock in front of it,
  shown in an HEE coordinate system.  Green, blue, and red arrows
  indicate the directions toward STEREO-A, Wind, and SolO,
  respectively.}
\end{figure}
     Trial and error combined with subjective judgment are
used to vary the parameters of the MFR and shock shapes until a
solution is found that is deemed to yield synthetic images that
best match the real ones.  The final morphological model is
shown in Figure~6.  The synthetic images associated with this
model are shown in Figures~3-4, for comparison with the real ones.
The two movies that accompany the figure provide a more
comprehensive comparison.  Because we only clearly see the shock
to the west of the ejecta in the coronagraphic images, our
reconstructed shock ends up asymmetric in that direction.
This could be misleading, as in the HI1-A data we actually do see
hints that the shock may extend eastward of the ejecta as well,
in a manner not reproduced by our reconstruction.  Regardless,
the shock is not a focal point of our study, as we are far more
interested in the MFR characteristics.

\begin{table}[t]
\small
\begin{center}
Table 1:  CME Morphological Parameters \\
\begin{tabular}{clcc} \hline \hline
Parameter & Description & MFR & Shock \\
\hline
$\lambda_s$ (deg)& Trajectory longitude &   2   &  12  \\
$\beta_s$ (deg)  & Trajectory latitude  &  -5   &  -5  \\
$\gamma_s$ (deg) & Tilt angle           &  30   & ...  \\
FWHM$_s$ (deg)   & Angular width        & 50.7  & 81.6 \\
$\Lambda_s$      & Aspect ratio         & 0.17  & ...  \\ 
$\eta_s$         & Ellipticity          & 1.6   & ...  \\
$\alpha_s$       & Leading Edge Shape   &   4   &   5  \\
$x_{max}$         & Leading Edge Dist.   & 1.0   & 1.05 \\
\hline
\end{tabular}
\end{center}
\end{table}
     Table~1 lists the best fit parameters of the 3-D reconstruction,
using the variable names from \citet{bew17}.  The central
trajectory is defined by $\lambda_s$ and $\beta_s$, in HEE
coordinates.  The MFR is directed only $\lambda_s=2^{\circ}$ west of
Earth in longitude and $\beta_s=-5^{\circ}$ south of Earth in latitude.
The $\gamma_s$ parameter is the tilt angle of the MFR, with
$\gamma_s=0^{\circ}$ corresponding to an E-W orientation parallel to
the ecliptic.  With a positive value of $\gamma_s$ ($\gamma_s=30^{\circ}$)
the west leg of the MFR is tilted upwards and the east leg downwards.
This tilt is clearly indicated by the images in Figures~3(b) and 4(a),
and it means that STEREO-A and Wind see a relatively direct impact of
the MFR, while SolO to the east encounters the MFR much farther
from its central axis.  We also note that this tilt actually
agrees well with the orientation of the flare in Figure~1.

     The $FWHM_s$ parameter is the full-width-at-half-maximum angular
width of the MFR, which we find to be $FWHM_s=50.7^{\circ}$.  The
aspect ratio, $\Lambda_s=0.17$, indicates the minor radius of the apex
of the MFR divided by the distance of the apex from the Sun.
Larger values correspond to fatter MFRs.  The assumed ellipticity of
the MFR channel is $\eta_s=1.6$, the major radius divided by the minor
radius.  This ellipticity is constrained by the shape of the
CME halo.  Reducing ellipticity and assuming a more
circular channel yields a rounder outline than observed in
Figures~3-4.  The $\alpha_s$ parameter defines the shape of the
MFR leading edge, with high values corresponding to flatter
leading edges.  The MFR shape and synthetic images of it
are not terribly sensitive to $\alpha_s$ in this case, but we make our
MFR flatter on top by assuming a relatively high value of $\alpha_s=4$.
Note that $\eta_s$ and $\alpha_s$ both affect time of arrival and
duration predictions for spacecraft encounters, and can be
modified to improve those predictions, if needed.  Finally,
the $x_{max}$ parameter exists simply to indicate how advanced beyond
the MFR the shock is presumed to be.

\section{Multi-Point Field Measurements of the CME}

     The primary reason that this CME is worthy of special attention is
that it hits multiple spacecraft near 1~au, leading to measurements
of field properties at multiple locations, thereby providing an ideal
opportunity to test the MFR paradigm for CME structure.  In this section,
we present the in~situ measurements of the CME from
Wind \citep{rpl95,kwo95}, STEREO-A \citep{mha08,abg08,jgl08},
and SolO \citep{tsh20,cjo20}.
These data are obtained from publicly accessible archives supported by
NASA (e.g., STEREO Science Center, SPDF/OMNI, SPDF/CDAWeb).

\begin{figure}[t]
\plotfiddle{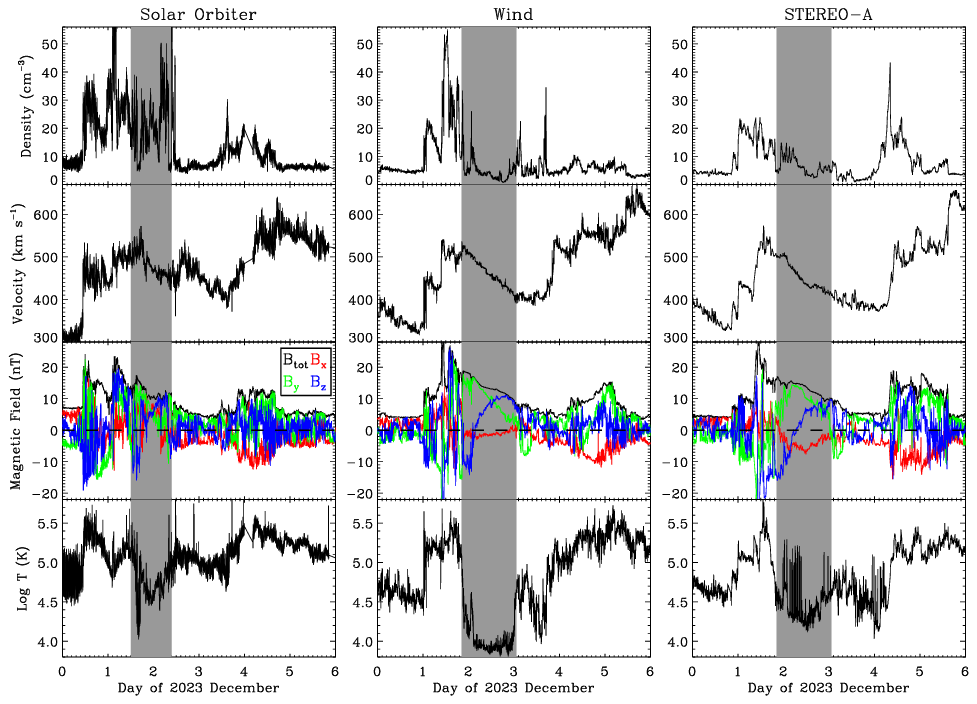}{3.3in}{0}{85}{85}{-270}{-340}
\caption{Plots of solar wind proton density, velocity, magnetic field,
  and temperature from 2023~November~30 through 2023~December~5, from
  Solar Orbiter, Wind, and STEREO-A.  The magnetic field panel shows
  the individual field components in an RTN coordinate system, in
  addition to the total field.  Shaded regions indicate the time periods
  that we associate with the MFR of the 2023~November~28 CME.}
\end{figure}
     Figure~7 provides an overview of the solar wind properties
observed by the three spacecraft between 2023~November~30 and
2023~December~5.  These data are interpreted using the MFR
encounter geometry implied by Figure~6, with the
shaded regions in Figure~7 indicating the time periods that
we associate with the MFR.  With the central CME trajectory being
only a little over $5^{\circ}$ away from Earth, mostly to the south,
it is Wind that is closest to the true apex of the MFR.
STEREO-A is nearer to the west leg, and SolO to the east leg.
The $\gamma_s=30^{\circ}$ tilt of the MFR centers the part of the MFR
that hits STEREO-A closer to the ecliptic plane, meaning that
STEREO-A experiences a near direct hit by the MFR, and actually comes
nearer to the central axis of the MFR than Wind does.  In contrast,
the MFR tilt shifts the part of the MFR that hits SolO away from
the ecliptic, meaning SolO experiences more of a grazing incidence
encounter with the northern edge of the MFR, significantly farther
from the MFR axis than either STEREO-A or Wind.  This leads to
a more muddled magnetic signature than is the case for STEREO-A
or Wind.  (A more direct comparison of the reconstructed MFR and the
in~situ data will be shown in the next section.)

     For Wind and STEREO-A, there is a clear phase of ordered, rotating
field from late on December~1 through December~2, with low temperature
and linearly decreasing velocity.  These are the kinds of signatures
often associated with clear in~situ CME encounters.  The decreasing
velocity is consistent with an MFR expanding in a roughly self-similar
fashion as it moves away from the Sun.  The field rotations are
very similar for Wind and STEREO-A, with $B_z$ rotating from negative
to positive values, and $B_y$ with generally decreasing positive
values.  Even though the MFR signatures for SolO are less clear and
ordered, it is important to note that the same $B_z$ and $B_y$ variation
is still apparent, as is the decreasing velocity.

     The MFR temperature signature shows interesting variation along
the structure.  The lowest plasma temperatures inside the MFR are seen
for Wind closest to the apex.  The temperatures at STEREO-A are
noticeably higher, and they are higher still at SolO.  The density
signatures show analogous variation, with the lowest internal MFR
densities seen by Wind, somewhat higher densities at STEREO-A, and
much higher densities at SolO.  The higher densities and temperatures
at SolO are consistent with the idea that regions of the MFR far from
its axis are more contaminated by incursions of plasma from
the ambient solar wind through which the MFR is traveling.  Such
incursions may be indicative of MFR erosion, possibly due to
small-scale reconnection events in the sheath region ahead of the MFR.
The observations of this CME could therefore provide useful
tests for models of CME MFR erosion.

     Signatures of the CME's interaction with the ambient solar wind are
also apparent in the extended sheath region ahead of what we are
interpreting as the MFR, which hits SolO midway through November~30 and
Wind/STEREO-A at the beginning of December~1, lasting roughly a day
in each case before the MFR arrival.  The sheath could be predominantly
quiescent solar wind heated and compressed by the CME shock.  With the
Sun being quite active, it could also include smaller eruptive magnetic
structures swept up by the faster CME between the Sun and 1~au.
\citet{yc24} interpret parts of the sheath as being due to two
particular CMEs, which are visible in coronagraph images early on
November~28, and are possibly swept up by the faster event studied here.

     It is also worth noting that after the MFR, a CIR structure sweeps
over the three spacecraft sequentially on December~3-4, from SolO to Wind
and finally to STEREO-A.  The CIR period is characterized by initial
enhancements in density and field strength, and by high wind velocites
and temperatures that persist past December~5.  This CIR is
caused by the high speed wind streams emanating from the coronal hole
apparent east of the flare in Figure~1.  In Section~3, we noted that
the HI2-A movies show visible fronts associated with this CIR.

\section{Field Insertion into the Model Flux Rope}

     In order to test the MFR paradigm, in this section we insert a
physically plausible field structure into the 3-D shape inferred from
white light imaging of the 2023~November~28 CME (see Figure~6),
in order to see if the resulting 3-D CME field model can simultaneously
reproduce the field signatures seen at all three spacecraft hit by the
CME near 1~au.  Our methodology for inserting a field structure into
our MFR shell is the same as that in \citet{bew20} for a
pair of CMEs from 2012~August~2 with both in~situ and radio Faraday
rotation constraints on field structure.  The physical basis for this
procedure is the MFR model of \citet{tnc18}, which
is notable for allowing for MFRs with elliptical cross sections.
Essentially, the \citet{tnc18} model is used to
define the 2-D field structure at the apex of our MFR, and then
simplifying physical assumptions are used to extrapolate that
apex field model throughout the 3-D MFR.

\begin{figure}[t]
\plotfiddle{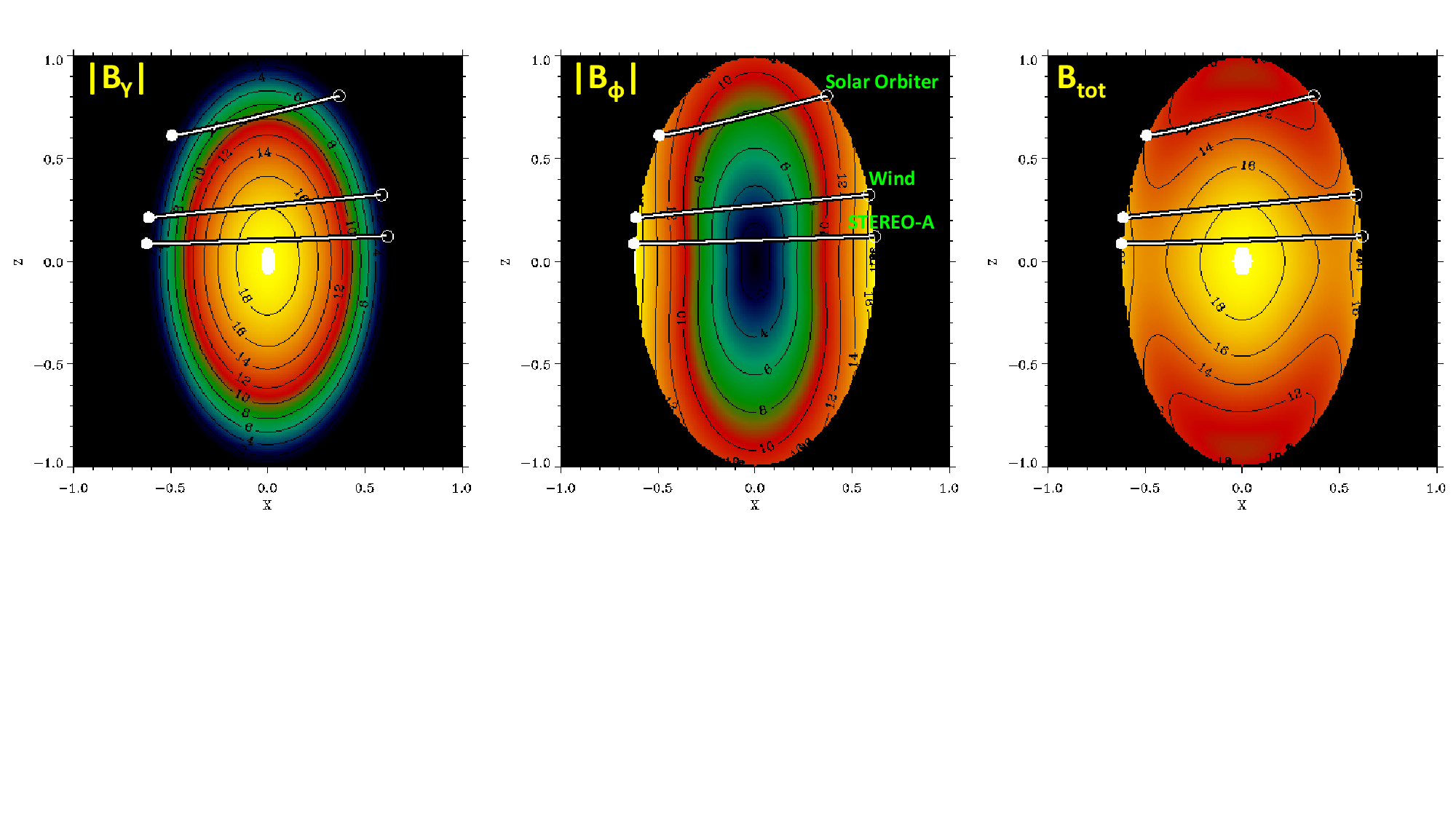}{2.3in}{0}{55}{55}{-270}{-120}
\caption{Maps of the magnetic field at the apex of the MFR for the
  2023~November~28 CME, based on the model that best fits the
  in-situ data from STEREO-A (see Figure~9).
  From left to right, the maps show the axial field ($|B_Y|$), the
  azimuthal field along the elliptical contours of the MFR channel
  ($|B_{\phi}|$), and the total field ($B_{tot}$).  The contours
  indicate the field values in units of nT.  The values shown
  correspond to the time when the leading edge of the CME
  reaches 1~au.  The white lines indicates the paths of Solar
  Orbiter, Wind, and STEREO-A through the MFR channel, which move
  from right to left through the MFR.}
\end{figure}
     With the ellipticity of the MFR having been determined
observationally by imaging, a 2-D apex field map for the MFR is defined
entirely by only two free parameters, the axial field at the MFR center,
$B_t$, and the maximum azimuthal field at the surface of the MFR, $B_p$.
We refer the reader to \citet{tnc18} and \citet{bew20}
for details of that computation.  Figure~8 shows
maps of axial field, $B_Y$, and azimuthal
field $B_{\phi}$, based on $B_t=-19.3$~nT and $B_p=+17.4$~nT,
best-fit values determined as described later.
The axial field has the peak value of $B_t$ at MFR center, decreasing
to zero at the MFR surface.  In contrast, $B_{\phi}$ is zero
at the axis but increases toward the MFR surface, with a maximum,
$B_p$, at the surface along the minor axis of the ellipse.  As a
sign convention, we assume the negative $B_t$ corresponds to a
direction into the plane for $B_Y$, and a positive $B_p$ corresponds
to a right-handed MFR, meaning the $B_{\phi}$ field lines follow
elliptical contours in a clockwise direction in Figure~8.

     The axial field is relatively easy to extrapolate from the apex
to anywhere within the MFR, since flux must be conserved along the MFR,
meaning that $B_Y$ must scale as the inverse of the cross-sectional
area along the MFR.  There is less clarity about what is best assumed
for $B_{\phi}$.  Following \citet{bew20}, we
assume $B_{\phi}\propto 1/a_{min}$, where $a_{min}$ is the local minor
radius of the MFR channel.  With this assumption, both $B_Y$ and
$B_{\phi}$ increase into the narrower legs of the MFR, but $B_Y$
increases more, making the field more axial in the legs than in the
apex.  It should be emphasized that a field structure inserted into
an arbitrary 3-D MFR shape in this fashion will not be divergence
free, and will only approximate a physically proper flux rope.

     The MFR fields are naturally time-dependent.  The reference time
that we use to define the initial field model, corresponding to the
apex maps in Figure~8, is when the leading edge distance from the Sun,
$R_{le}$, of the MFR is 1~au.  Because we are assuming simple
self-similar expansion of the MFR with time, this implies $B_Y$ and
$B_{\phi}$ both scale as $1/R_{le}^2$.  The kinematic model in
Figure~5, specifically the top panel, explicitly shows how $R_{le}$
changes with time.  With this information, we can now take an apex
field map like that in Figure~8 and not only extrapolate it anyplace
within the MFR, but also extrapolate it to any time during the MFR's
expansion into interplanetary space.

     With the 3-D field insertion procedure now described, we can
now focus on the field properties predicted by this model for the
three spacecraft hit by the CME.  Figure~8 provides an indication
of the tracks of the spacecraft through the MFR channel, with the
spacecraft moving from right to left through the channel.  This
depicts more explicitly the encounter geometries described
in Section~4.  Thanks to the $\gamma_s=30^{\circ}$ tilt of the MFR,
STEREO-A near the west leg of the MFR sees a more direct hit than
Wind, coming closer to the central axis, even though Wind is closer
to the apex.  In contrast, the $\gamma_s=30^{\circ}$ tilt shifts SolO
near the east leg of the MFR away from its axis.
The $+3.8^{\circ}$ north latitude of SolO above the ecliptic also
contributes to this shift.  Incidentally, the reason the spacecraft
tracks in Figure~8 show positive slopes is because of the presumed
self-similar expansion of the MFR as it encounters the spacecraft.

\begin{figure}[t]
\plotfiddle{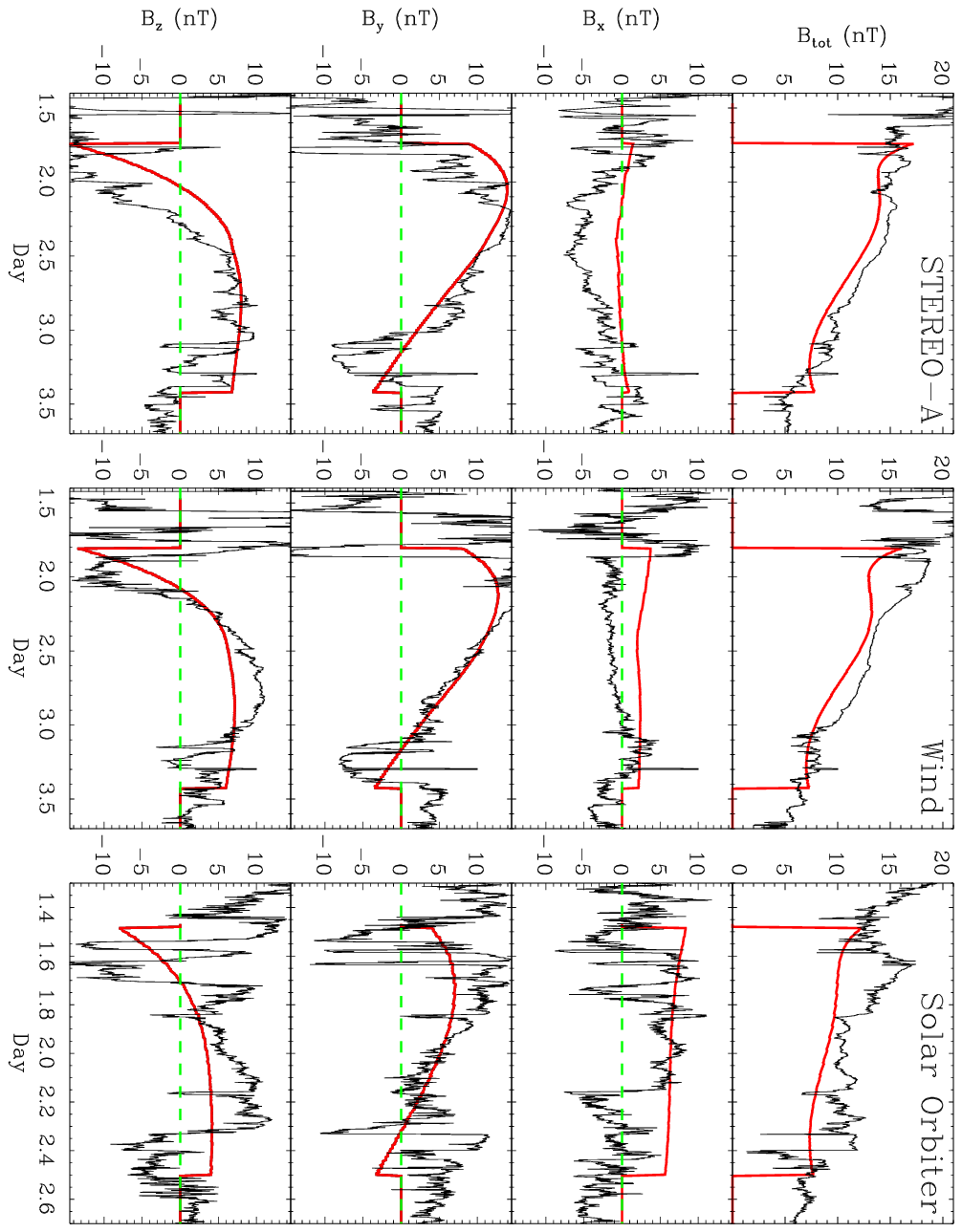}{3.7in}{90}{65}{65}{260}{-65}
\caption{Plots of $B_{tot}$, $B_x$, $B_y$, and $B_z$ observed by
  STEREO-A, Wind, and Solar Orbiter in 2023~December, showing the
  field variations associated with the MFR of the 2023~November~28
  CME.  The STEREO-A data are fitted with a field model inserted
  into the MFR structure shown in Figure~6.  Red lines show the
  resulting best fit with $B_t=-19.3$~nT and $B_p=+17.4$~nT,
  shifted forward by 0.65 days.  This
  model is then used to predict the field that Wind and Solar Orbiter
  should see, shown as red lines in those panels.}
\end{figure}
     Figure~9 shows the field components observed by the three
spacecraft as they encounter the MFR.  These can be compared with
MFR model predictions, but first an assessement must be made about
the degree to which the kinematic and morphological reconstruction
described in Sections~3 and 4 successfully predict the MFR arrival
times and encounter durations at the three spacecraft.  We find that
the arrival time predictions are very poor, but the encounter duration
predictions are acceptable.  In comparing model predictions with
data, we have to arbitrarily shift the arrival time forward
by 0.65~days.  The shortcomings of the kinematic model have already
been noted in Section~3.  The early arrival time prediction of the
model exists despite a final velocity prediction, 558 km~s$^{-1}$ (see
Figure~5), that is somewhat faster than the observed peak
velocity, $\sim 520$ km~s$^{-1}$ (see Figure~7).

     In fitting the in~situ field data to find the best MFR model,
we decide to fit only the STEREO-A data, which show the closest
passage to the MFR axis, and we then see if the resulting
MFR model can successfully reproduce the field behavior seen at
Wind and SolO.  We simultaneously fit all four of the field
quantities shown in Figure~9 ($B_{tot}$, $B_x$, $B_y$, and $B_z$).
Fitting $B_{tot}$ seems redundant, but we believe it is advantageous
for the fit to include this positive-definite quantity as a
constraint.  As noted above, there are actually only two free
parameters of this fit, $B_t$ and $B_p$.  The best least-squares
fit to the STEREO-A data is shown in Figure~9, with $B_t=-19.3$~nT
and $B_p=+17.4$~nT, with the apex field maps shown in Figure~8.
The negative $B_t$ value means that the left (e.g., east) leg of
the MFR in Figure~6 is the positive polarity leg, and the right
(e.g., west) leg is negative polarity.  Incidentally, we find that
a fit to the Wind data instead of STEREO-A leads to a similar
quantitative result, with $B_t=-21.9$~nT
and $B_p=+18.1$~nT.  The STEREO-A MFR model is used to
predict the field tracings at Wind and SolO, and these predictions
are also compared with the Wind and SolO data in Figure~9.

     The fit is reasonably successful at reproducing the observed
field signs and rotations, especially considering that there are
only two free parameters.  The STEREO-A fit predicts the Wind
field measurements well, as expected given that the paths of
STEREO-A and Wind through the MFR channel are similar and the
observed field orientations are as well.  There are still small
but clear differences in field between Wind and STEREO-A despite
the small $6.5^{\circ}$ spacecraft separation, but field differences
have been seen for even smaller separations \citep{fr24}.
The factor of two
decrease in total field strength during the STEREO-A and Wind MFR
encounters is largely due to the self-similar expansion of the MFR
during the encounters.  The change in $B_z$ sign from negative to
positive indicates a clockwise azimuthal $B_{\phi}$ field,
corresponding to the positive $B_p$.  The $B_x$ components observed
for STEREO-A and Wind are lower than predicted by the model by
about 5~nT.  The simplest way to explain this is if the front of
the real CME MFR is not perfectly perpendicular to the CME
trajectory direction, but has a $\sim 15^{\circ}$ tilt in the
right direction for the axial field, $B_Y$, to have a small
negative $B_x$ component.

     The best test of the MFR field model comes from the SolO
measurements, given that SolO is farther from STEREO-A than Wind,
and encounters the MFR much farther from its central axis
than either STEREO-A or Wind.  In a broad sense, the model successfully
predicts the general field signs and rotations observed by SolO.
The $B_y$ component is positive but decreasing,
as predicted, and $B_z$ generally changes from negative to
positive, also as predicted.

      However, it is $B_x$ that is particularly worthy of note.
The $B_x$ component is mostly from the azimuthal field, $B_{\phi}$,
with the x-direction pointing roughly to the right in the $B_{\phi}$
panel of Figure~8.  With STEREO-A and Wind passing close to the MFR
axis, they are not predicted to see much field in the $B_x$
direction, and they don't.  In contrast, SolO grazes the top of the
MFR where the azimuthal field is partly in the positive x-direction.
Thus, SolO is predicted to see a significant positive and roughly
flat $B_x$ field profile, and this is in fact observed.  This is
the most impressive successful prediction of the MFR model.  We
conclude that a simple MFR model with an orientation inferred
from imaging data is able to simultaneously explain
the field tracings seen by the three spacecraft that encounter the
2023~November~28 CME to an acceptable degree, and that these data
therefore provide support for the MFR paradigm.

     However, the SolO field measurements within the MFR are not
nearly as smooth as for STEREO-A and Wind, showing far more
excursions and anomalies.  We previously noted in Section~5 the
higher plasma densities and temperatures seen within the MFR region
by SolO.  We interpret these measurements as indications of the
erosion and degradation of the MFR along its periphery.  The
observations of this CME are particularly useful for studying this,
since there are measurements of both the periphery of the MFR
channel, from SolO, and measurements near the MFR axis for
comparison, from Wind and STEREO-A, showing a much cleaner
MFR signature.

\section{Summary}

     We have presented both imaging and in~situ observations of an
Earth-directed CME that erupted on 2023~November~28.  The CME is of
particular interest for hitting three spacecraft near 1~au:  Wind
near Earth, STEREO-A $6.5^{\circ}$ west of Earth, and
SolO $10.7^{\circ}$ east of Earth in longitude.
Our findings can be summarized as follows:
\begin{enumerate}
\item We perform a full 3-D morphological and kinematic
  reconstruction of the event, based on imaging from SOHO/LASCO
  and STEREO-A.  This yields an MFR shape for the CME, which is
  assumed to expand away from the Sun in a self-similar manner,
  with a peak speed of 1270 km~s$^{-1}$, decelerating to a final
  speed of 558 km~s$^{-1}$.
\item Wind and STEREO-A are inferred to sustain direct hits with
  the MFR, passing close to its central axis.  In contrast, SolO
  experiences a more grazing incidence encounter along the
  northern edge of the MFR.  The MFR signatures in the Wind and
  STEREO-A in~situ data are fairly clear, with smoothly rotating
  fields, linearly declining velocity, low temperature, and low
  density.  In contrast, the SolO MFR signature is more confused,
  with higher temperatures, much higher densities, and numerous
  anomalies in the field measurements.  The interpretation is
  that this is due to substantial MFR erosion and degradation
  on the periphery of the MFR.
\item With no lateral view of the CME, our kinematic model
  of the event carries large uncertainties, leading to inaccurate
  predictions for the CME's arrival at Wind, STEREO-A, and SolO.
  We have to shift the arrival time forward by 0.65 days to agree
  with the in~situ data.
\item For confronting the in~situ data, we insert a plausible
  field structure into the MFR shape based on the methodology
  described by \citet{bew20} and \citet{tnc18},
  which allows the field structure to be entirely
  defined by two parameters:  the central axial field at the MFR
  apex, $B_t$, and the peak azimuthal field at the surface of the
  MFR at its apex, $B_p$.  We first fit the STEREO-A in~situ data,
  finding that $B_t=-19.3$~nT and $B_p=+17.4$~nT best fit these
  data.  We then assess whether this MFR model also fits the Wind and
  SolO in~situ data, and we find that it does so reasonably well.
\item The general agreement with the SolO data means that we have
  demonstrated that for this event our MFR field model can
  successfully be used to infer field measurements made along the
  periphery of the MFR in one location (e.g., SolO) based on field
  measurements made $17^{\circ}$ degree away near the MFR axis
  by another spacecraft (e.g., STEREO-A).  This provides
  support for the MFR paradigm for CME field structure, and
  demonstrates that despite evidence of MFR erosion and
  degradation on the periphery of the MFR, there is still a
  coherent large-scale MFR field structure defining the core
  of this CME.
\end{enumerate}

\acknowledgments

Financial support was provided by the Office of Naval Research.
We acknowledge use of NASA/GSFC's Space Physics Data Facility's
OMNIWeb and CDAWeb services, and OMNI data.  We also acknowledge the
use of images from the SDO mission, courtesy of the NASA SDO and AIA
science teams.  The STEREO imaging data from the SECCHI instrument
are produced by a consortium of NRL (US), LMSAL (US), NASA/GSFC (US),
RAL (UK), UBHAM (UK), MPS (Germany), CSL (Belgium), IOTA (France),
and IAS (France).  In addition to funding by NASA, NRL also
received support from the USAF Space Test Program and ONR.
This work has also made use of data provided by the STEREO PLASTIC
and IMPACT teams, available at the STEREO Science Center website,
supported by NASA contracts NAS5-00132 and NAS5-00133.
Solar Orbiter is a mission of international cooperation between ESA
and NASA, operated by ESA.  Solar Orbiter magnetometer (MAG) data was
provided by Imperial College London and supported by the UK Space
Agency.  The Solar Orbiter Solar Wind Analyser (SWA) scientific
sensors, SWA-EAS, SWA-PAS, SWA-HIS, and the SWA-DPU have been
designed and created, and are operated under funding provided in
numerous contracts from the UK Space Agency (UKSA), the UK Science
and Technology Facilities Council (STFC), the Agenzia Spaziale
Italiana (ASI), the Centre National d’Etudes Spatiales
(CNES, France), the Centre National de la Recherche Scientifique
(CNRS, France), the Czech contribution to the ESA PRODEX programme,
and NASA.

\end{document}